## Multilevel models for continuous outcomes

George Leckie

Centre for Multilevel Modelling and School of Education, University of Bristol

**Address for correspondence**

Centre for Multilevel Modelling

School of Education

University of Bristol

35 Berkeley Square

Bristol

BS8 1JA

United Kingdom

g.leckie@bristol.ac.uk

## Multilevel models for continuous outcomes

**Abstract**

Multilevel models (mixed-effect models or hierarchical linear models) are now a standard approach to analysing clustered and longitudinal data in the social, behavioural and medical sciences. This review article focuses on multilevel linear regression models for continuous responses (outcomes or dependent variables). These models can be viewed as an extension of conventional linear regression models to account for and learn from the clustering in the data. Common clustered applications include studies of school effects on student achievement, hospital effects on patient health, and neighbourhood effects on respondent attitudes. In all these examples, multilevel models allow one to study how the regression relationships vary across clusters, to identify those cluster characteristics which predict such variation, to disentangle social processes operating at different levels of analysis, and to make cluster-specific predictions. Common longitudinal applications include studies of growth curves of individual height and weight and developmental trajectories of individual behaviours. In these examples, multilevel models allow one to describe and explain variation in growth rates and to simultaneously explore predictors of both of intra- and inter-individual variation. This article introduces and illustrates this powerful class of model. We start by focusing on the most commonly applied two-level random-intercept and -slope models. We illustrate through two detailed examples how these models can be applied to both clustered and longitudinal data and in both observational and experimental settings. We then review more flexible three-level, cross-classified, multiple membership and multivariate response models. We end by recommending a range of further reading on all these topics.

## 1. Introduction

Multilevel models (mixed-effect models or hierarchical linear models) are now a standard extension to conventional regression models for analysing clustered and longitudinal data and are widely applied in the social, behavioural and medical sciences (Goldstein, 2011; Fitzmaurice et al., 2011; Hedeker, D. & Gibbons 2006; Hox et al., 2017; Raudenbush and Bryk, 2002; Skrondal and Rabe-Hesketh, 2004; Singer and Willett, 2003; Snijders and Bosker, 2012). In this article we focus on multilevel linear regression models for continuous responses (outcomes or dependent variables).

Just as with conventional linear regression, the purpose of multilevel linear regression is to model the relationships between a continuous response and a set of covariates (predictors or explanatory variables). A key difference is that the covariates may be defined at different levels of analysis. For example, when the data are clustered (e.g. data on student achievement test scores across many schools) the covariates may be at the individual-level (student) or at the cluster-level (school). Furthermore, the regression relationships between lower-level covariates and the response may differ at different levels of analysis and it is often important to disentangle these within- and between-cluster effects and to interpret them separately. Importantly, even after adjusting for the covariates, unobserved heterogeneity typically remains across clusters (due to omitted cluster-level variables) which induces a clustering (dependency, correlation or similarity) in the observed responses. This violates the linear regression assumption of independent residuals. The data contain less information than naively assumed by linear regression. This leads to incorrect standard errors which are typically too small, especially for higher-level covariates. Thus, in the presence of clustering, conventional linear regression may result in spuriously precise parameter estimates and type I errors of inference. In addition, the cluster effects themselves are often of substantive interest as they allow one to make comparisons across clusters which have been adjusted for the mix of individuals in each cluster. As such they are widely used in the production of institution league tables in education, health and other areas of application. More generally, multilevel modelling allows one to study how the regression relationships vary across clusters and to identify those cluster characteristics which predict such variation.

Common examples of clustered data include data on students nested in schools, patients nested in hospitals, respondents nested in neighbourhoods, and employees nested in firms. In each case we have two-level data with individuals (level-1) nested within clusters (level-2). In some applications these clusters may themselves be nested within superclusters leading to three-level data. For example, students (level-1) nested in schools (level-2) nested in school districts (level-3). Multilevel linear regression extends naturally to incorporate and study this additional complexity. Sometimes there is no strict hierarchy and the level-1 units are best viewed as being separately nested within two or more different classifications of clusters but where those classifications are themselves not nested within one another. For example students (level-1) are nested within neighbourhoods (level-2) but not all students from the same neighbourhood attend the same school and so the data are not a strict three-level hierarchy with schools at level-3, rather the data are said to be cross-classified and neighbourhoods and schools are both viewed conceptually at level-2.

While longitudinal data are quite different from clustered cross-sectional data, such data can also be viewed as clustered. Common examples include repeated measurements of subjects, or panel waves nested within individuals. In each case we have two-level data with the longitudinal measurements (level-1) nested within individuals (level-2). In some applications these individuals may be further nested within, for example, organisations or areas leading to three-level data. In some cases, individuals will additionally change organisations or move areas from one occasion to the next, again violating the assumption of a three-level hierarchy. The resulting data are then again cross-classified but this time with individuals and their contexts both viewed conceptually at level-2.

It follows that similar data complexities and modelling issues arise when studying longitudinal data as arise when studying clustered cross-sectional data and we can use the same kinds of multilevel models to address them. Thus, with longitudinal data we tend to again have covariates defined at different levels of analysis (time-varying and time-invariant covariates) and interest lies in studying both inter- and intra-individual variation in the response simultaneously. Typically, there will also be substantial unobserved individual heterogeneity and in many applications there is also interest in allowing for individual specific time trends or trajectories (e.g., developmental trajectories or growth curves) and in then studying how these vary across individuals and by their characteristics.

In this article we shall introduce the most commonly applied multilevel models for continuous responses: two-level variance-component, random-intercept, and random-slope models. We shall show how all these models can be applied to both clustered and longitudinal data. We will illustrate these models and concepts in the context of two different applications in educational research, one based on observational data, one based on experimental data. The data and Stata software syntax to allow the reader to replicate all our results are available from the author upon request. Finally, we discuss additional reading including classic textbooks on multilevel modelling, software, and specific references to the literature on the most common modelling extensions beyond those considered here, including three-level and cross-classified models.

## 2. Two-level models for clustered data

In this section we introduce two-level variance-component, and random-intercept models for clustered data. We delay the introduction of random-slope models to the next section.

**Inner-London schools exam scores**

We shall illustrate multilevel modelling of clustered data using the inner-London schools exam scores data first analysed by Goldstein et al. (1993). These data are also used as the main example in the MLwiN multilevel modelling software user manual (Charlton et al., 2019; Rasbash et al., 2019). The data contain student exam scores from a number of schools as well as various student and school background characteristics. They are an example of the type of student-level data used in many education systems around the world to produce school league tables to hold schools to account and to guide parents choosing schools for their children. At its simplest this involves comparing school mean exam scores. A fundamental limitation of this approach is that schools differ greatly in their student composition at intake, with students in some schools academically much further ahead than students in other schools. One of the aims of the original study was to therefore promote a 'value-added' approach to school comparisons which adjusts school mean exam scores for students' achievements at entry and to therefore lead to fairer and more meaningful comparisons between schools. The authors used two-level models to make these adjustments and to produce these school specific estimates. This approach is now widely applied and recent discussions can be found in Leckie and Goldstein (2009, 2017, 2019). We will carry out a similar analysis here.

The data consist of 4,059 students (level-1) nested within 65 schools (level-2). The continuous response variable is an exam score based on students' age 16 general certificate of secondary education exam results (national exams in England taken at the end of compulsory secondary schooling). The key covariate is student age 11 prior achievement in the London reading test. We refer to these scores as students' age 16 and age 11 scores respectively. Both scores have been standardised to have zero mean and variance one to facilitate interpretation. The original scales are not well known to non-specialists whereas with standardised scores a 1-unit difference on each variable corresponds to a 1 SD difference. Other covariates we will consider include student gender (1623 boys; 2436 girls) and school gender (35 mixed schools; 10 boys' schools; 20 girls' schools).

**Model 1: Estimating the degree of clustering in student age 16 scores**

Before we start any modelling, we graphically inspect the data. Figure 1 plots student and school mean age 16 scores on the vertical axis against school on the horizontal axis. The plot shows that mean age 16 scores vary between schools; some schools score, on average, higher than others. This implies that students who attend the same school are more alike (positively correlated) in their age 16 scores than students who attend different schools. The data therefore appear clustered. We can also see that age 16 scores vary considerably within schools and indeed that the variation within schools appears to dominate the variation between schools.

We will start by fitting the simplest possible two-level model to these data: a two-level variance components model. This model includes no covariates (it is an unconditional, null or empty model); it simply allows one to formalise the patterns seen in Figure 1 by estimating and statistically testing the degree of clustering in student age 16 scores. Let $y_{ij}$ denote the age 16 score for student $i$ ($i = 1, \ldots, n_j$) in school $j$ ($j = 1, \ldots, 65$). (While the use of $i$ to denote the level-1 units and $j$ to denote the level-2 units is standard in social science, we note that in some areas of biostatistics the indices are sometimes used the other way around.) The model can then be written as

$$y_{ij} = \beta_0 + u_j + e_{ij} \qquad (1)$$

where $\beta_0$ denotes the intercept, $u_j$ denotes the school random intercept effect, and $e_{ij}$ denotes the student residual.

The intercept $\beta_0$ measures the average age 16 score across all schools and all students; it estimates the London-wide average age 16 score. Thus, this intercept measures the overall average of the data plotted in Figure 1. The summation $\beta_0 + u_j$ measures the average age 16 score in school $j$ and so $u_j$ measures the difference between school $j$'s performance and the overall average. These random effects capture the vertical variability in the school means plotted in Figure 1. The residual $e_{ij}$ measures the difference between each student's actual age 16 score and the score predicted by the school they attend. These residuals capture the variability of the student scores around their school means in Figure 1.

The school random intercept effects are assumed normally distributed with zero mean and between-school variance $\sigma_u^2$ while the student residuals are assumed normally distributed with zero mean and within-school variance $\sigma_e^2$.

$$u_j \sim N(0, \sigma_u^2) \qquad (2)$$

$$e_{ij} \sim N(0, \sigma_e^2) \qquad (3)$$

We fit this and subsequent models by maximum likelihood estimation, the standard method for fitting multilevel models. Table 1 presents the model results. The intercept $\beta_0$ is estimated to be -0.013 effectively equal to the sample mean of 0 (recall that student age 16 scores are standardised to have zero mean). The between school-variance $\sigma_u^2$ is estimated to be 0.169. One way to interpret the magnitude of this parameter is to calculate the range of scores within which we expect to find the middle 95% of schools in the population ($\beta_0 - 1.96\sigma_u$, $\beta_0 + 1.96\sigma_u$). We obtain (-0.819,0.793). Thus, students in the highest scoring schools (schools at the 97.5[th] percentile) are on average predicted to score 1.6 SD higher than students in the lowest scoring schools (schools at the 2.5[th] percentile). There are clearly substantial and meaningful differences in mean exam results between schools. The sample data plotted in Figure 1 are consistent with this prediction: the vast majoring of schools lie within this predicted range. The within-school variance $\sigma_e^2$ is estimated to be 0.848 and, as suggested by Figure 1, greatly exceeds the between-school variance. The total variance $\sigma_u^2 + \sigma_e^2$ is estimated to be 1.017 effectively equal to the sample variance of 1 (recall that student age 16 scores are standardised to have variance one).

The presence of the school random intercept effects capture the mean differences between schools. Their introduction implicitly allows for a positive correlation between students within the same school. In contrast, students from different schools continue to be assumed independent, just as they are in conventional linear regression. It can be shown that the correlation between two students from the same school is given by

$$\rho \equiv \text{Corr}(y_{ij}, y_{i'j}) = \frac{\sigma_u^2}{\sigma_u^2 + \sigma_e^2} \tag{4}$$

This correlation, which can range from 0 (no correlation) to 1 (perfect positive correlation), is referred to as the intraclass correlation coefficient (ICC). In our case the ICC is estimated to be 0.166 which is a modest correlation and inline with estimates found in the literature.

Examining the ICC expression, we see that it is calculated as the ratio of the between-school variance $\sigma_u^2$ to the total variance $\sigma_u^2 + \sigma_e^2$. For this reason, the ICC is also often interpreted as the proportion of response variation which lies between schools. When this second interpretation is used, the ICC is sometimes referred to as a variance partition coefficient (VPC). Thus, we can also state that 16.6% of age 16 score variation lies between schools, 83.4% within schools.

While the degree of clustering in the data suggested by Figure 1 and now estimated by the model appears substantively meaningful, it is nevertheless prudent to check that this result is also statistically significant. The relevant test is a test of the null hypothesis $H_0: \sigma_u^2 = 0$ against the alternate hypothesis $H_1: \sigma_u^2 > 0$. A $z$-ratio for the between-school variance greatly exceeds 2 and so it seems clear that the clustering is significant. However, z-tests of the variance components (and Wald tests more generally) are only approximate as they assume the parameters have normal sampling distributions when this is not the case (the cluster variance exhibit positive skew especially when there are only a limited number of clusters). The preferred way to test the significance of $\sigma_u^2$ is to therefore perform a likelihood-ratio test of the current model versus a simpler model where we omit the school random intercept effect (i.e., a linear regression with no covariates, not shown). The likelihood-ratio test statistic is then the difference in deviance statistics between the two models. (The deviance statistic is calculated as minus two times the log-likelihood statistic and so the lower the deviance, the better the model fit.)

$$L = D_1 - D_2 = 11509 - 11011 = 498 \tag{5}$$

In our case $L = 498$ which greatly exceeds the critical value of 3.84 for testing at the 5% level. The resulting $p$-value is tiny $p < 0.001$. The school effects are statistically significant and so the observed clustering is statistically as well as substantively significant and a two-level analysis is preferred to a conventional linear regression analysis. (We note that this $p$-value is conservative and strictly it should be divided by 2 to reflect that we are testing on the boundary of the parameter space, but this makes little difference here).

Post-estimation, it is often of interest to calculate and inspect the predicted random effects and residuals to assess the plausibility of the distributional assumptions made for these quantities. Thus, here we might plot histograms or quantile-quantile plots to assess the degree to which the random effects and residuals are normally distributed. In later models with covariates, we might additionally plot the predicted random effects and residuals against the individual covariates to inspect the degree to which the random effects and residuals are each homoscedastic (assumptions of the model). In many applications it will also be of interest to calculate and inspect the predicted random effects because they are substantively interesting in their own right as will be the case here. However, we shall delay doing this until the next model where we seek to interpret the predicted school effects as more meaningful value-added school effects.

## Model 2: Adding student age 11 scores and predicting the school effects

We now add student age 11 scores to the model. We refer to the resulting model as a random-intercept model. We reserve the term variance-components model for models with no covariates.

To motivate adjusting for student age 11 scores, the left panel of Figure 2 plots student age 16 scores against student age 11 scores. As one would expect, there is a strong predictive relationship between the two variables ($r = 0.59$). Crucially, the right panel of Figure 2 shows that a similar strong positive relationship plays out at the school level ($r = 0.69$). Thus, the schools with the highest mean age 16 scores tend to be the schools which had the highest achieving students at intake. It therefore does not make sense to attribute the differences in school average age 16 scores plotted in Figure 1 solely to differences in the effectiveness of these schools as a substantial proportion of these differences clearly reflects differences in student achievement that were present at intake and therefore outside the control of the school. We therefore enter student age 11 scores into the model to attempt to adjust the predicted school effects for these initial differences.

The two-level random-intercept model for these data can be written as

$$y_{ij} = \beta_0 + \beta_1 x_{1ij} + u_j + e_{ij} \qquad (6)$$

where $x_{1ij}$ denotes the age 11 score and $\beta_1$ is the associated regression coefficient. As with conventional linear regression, this model captures the overall average linear relationship in the data $\beta_0 + \beta_1 x_{1ij}$ but in addition it also captures 65 school-specific linear relationships $\beta_0 + \beta_1 x_{1ij} + u_j$. Crucially, these school regression lines are just intercept shifts of the average regression line. It may help to look ahead to the left panel of Figure 3 which plots the model predictions for these school lines.

Table 1 presents the model results. The intercept $\beta_0$ and slope $\beta_1$ of the overall average line are estimated to be 0.002 and 0.563. This is the black line in the left panel of Figure 3. Thus, in the average school, a student with an age 11 score of 0 (i.e., an average student) is predicted to score 0.002 at age 16 while a 1-unit increase in student age 11 score (i.e., a 1 SD increase) is predicted to increase student age 16 score by 0.563 units. The slope coefficient has a z-ratio of $z = \hat{\beta}_1 / \text{SE}(\hat{\beta}_1) = 47$, far in excess of the critical value of 1.96 for testing at the 5% level. Student age 11 scores are clearly a significantly predictor of age 16 scores ($p < 0.001$). (In contrast to variance parameters, z-ratios and Wald tests more generally are valid for testing the regression coefficients).

Including student age 11 scores in the model leads the estimated total variance to reduce from 1.017 to 0.658. The percentage reduction is 35% ($= 100(0.658 - 1.017)/1.017$) and this latter statistic may be reported as the overall R-squared statistic for the current model. We can also report level-specific R-squared statistics. The estimated between-school variance has reduced from 0.169 to 0.092 and so 45% of the variation in schools' age 16 performances is explained by school-level differences in their age 11 scores.

Thus, schools are clearly a far less important determinant of students' age 16 scores than naively implied by Figure 1. However, we shall see that the remaining school differences are still sizeable. The within-school variance has reduced from 0.848 to 0.566 and so 33% of the within-school variation in students' age 16 scores is explained by variation in students' age 11 scores. Interestingly, the explanatory power of student age 11 scores is therefore greater at the school level than at the student level, again highlighting the substantial school differences in student achievement at intake.

Having added a covariate to the model, it is always sensible to recalculate the ICC/VPC and this can be done using the same expressions as before. This statistic now measures the residual clustering in the data (the clustering in the response having adjusted for the covariate). We obtain an estimate of 0.140 and so 14% of the variation in adjusted age 16 scores (progress, improvement or learning between the start and end of secondary schooling) lies between schools as opposed to 0.166 or 17% of the variation in unadjusted age 16 scores (final achievement at the end of secondary schooling). Thus, student progress varies less across schools than does student final achievement. A likelihood-ratio test confirms that this now lower degree of clustering is still statistically significant and so a multilevel model is still required, even after adjusting for student age 11 scores ($L = 403$, $p < 0.001$).

The school random-intercept effects $u_j$ can now be considered as 'value-added' school effects which measure the effect each school has on student age 16 scores having controlled for student achievement at intake. Post-estimation we can assign values to these effects in order to examine the effectiveness of each school. The standard approach is to use empirical Bayes prediction (shrinkage estimation or best linear unbiased prediction). An interesting feature of such predictions is that they exhibit 'shrinkage'. Essentially, the predicted school effects are shrunk towards the overall average of zero with greater shrinkage exhibited for schools with the fewest students. Shrinkage is desirable, both here and more generally, because it only affects schools that provide little information and it effectively downplays their influence, borrowing strength from other schools. All else equal, in absence of shrinkage, the concern is that small schools would disproportionally appear to be the least and most effective schools in the sample simply by virtue of chance variation. Shrinkage also leads to narrower 95% confidence intervals. Thus, by borrowing strength we obtain not just more reasonable predictions, but we are able to state these with more precision than if we ignore the fact that the schools come from a common distribution.

The left panel of Figure 3 plots the predicted mean relationship and the 65 predicted school lines. The plot shows some schools lines are located much higher on the plot than others suggesting that those schools impart positive effects on their students relative to those schools at the bottom of the plot which impart negative effects. However, what the plot fails to communicate is the extent to which these predicted relative school effects can be statistically separated. Figure 4 attempts to address this limitation by presenting a graphical league table of the predicted school effects together with their 95% confidence intervals. This plot is sometimes referred to as a 'caterpillar plot'. The 95% confidence intervals for many schools overlap and so the predicted school effects are clearly rather imprecise. The middle half of schools (34 schools) cannot be statistically separated from the average school.

To see the benefit of shrinkage consider school 54. This school has an estimated effect of -0.547 and is ranked 3rd from bottom of the league table. It is predicted to be one of the least effective schools in the sample. However, in absence of shrinkage, the estimated effect would take a much lower value of -0.963 and the school would be ranked bottom. However, we only observe eight students in this school and so the random variation (chance, noise) component in this schools' performance is so high that shrinkage plays a useful role preventing users from potentially overinterpreting what are rather noisy data.

**Model 3: Adding further student- and school-level covariates**

Having predicted and described the value-added school effects, interest often turns to attempting to explain why some schools appear more effective than others. Similarly, there is interest in using these models to explain why, within any given school, some students progress more rapidly than others. We can address such questions by adding further student and school covariates to the model. Here we explore the role of student and school gender. The extended model can be written as

$$y_{ij} = \beta_0 + \beta_1 x_{1ij} + \beta_2 x_{2ij} + \beta_3 x_{3j} + \beta_4 x_{4j} + u_j + e_{ij} \qquad (7)$$

where we enter student gender into the models as a female dummy variable $x_{2ij}$ and where we enter the nominal three-category school gender variable as two dummy variables $x_{3j}$ and $x_{4j}$ for attending an all boys' school or an all girls' school. The reference or omitted category is attending a mixed-sex school.

Table 1 presents the results. Girls in mixed-sex schools make 0.167 ($z = 4.91$, $p < 0.001$) of a SD more progress than boys in mixed-sex schools. Boys in all boys' schools make 0.178 ($z = 1.60$, $p = 0.109$) of a SD more progress than boys in mixed-sex schools. Girls in all girls' schools make 0.159 ($z = 1.82$, $p = 0.068$) of a SD more progress than girls in mixed-sex schools. Thus, there appears to be an advantage associated with attending single sex schools, but neither of the single-sex school effects are statistically significant at the 5% level. ($\beta_4$ is estimated more precisely than $\beta_3$ as there are 20 all girls' schools vs. 10 all boys' schools).

For pedagogic purposes, it is instructive at this point to answer the question: What would happen if we ignore the school-level clustering? To answer this, we fit a conventional linear regression version of the previous model (results not shown). Boys now score 0.183

SD higher in boys vs. mixed schools ($z = 4.29$, $p = 0.003$). Girls now score 0.168 SD higher in girls vs. mixed schools ($z = 5.16, p < 0.001$). Now both single-sex school effects appear highly significant, even at the 1% level, but these inferences are not valid as we have incorrectly assumed the residuals to be independent when we know from Model 3 there is substantial residual clustering in the data (ICC/VPC = 0.130). This is an illustration of the general result that when one ignores clustering one obtains spuriously precise estimates for the regression coefficients, especially for cluster level covariates. Ignore clustering at your peril.

**Further extensions**

Perhaps the most common extension to the two-level models presented here is to allow not just the intercept to vary across clusters, but to also allow one or more of the slope coefficients to vary across clusters. This would allow the regression relationships to vary from cluster to cluster. In the context of the application presented here, this would allow the 65 school lines to potentially crossover one another and this is illustrated in the right panel of Figure 3 (these lines are from a random-slope version of Model 2). This raises interesting and important questions since the school which appears most effective for students with high age 11 scores is no longer the same as the school which appears most effective for students with low age 11 scores. We introduce random-slope models in the next section in the context of an application to longitudinal data.

Another important extension to consider in two-level models is whether one needs to decompose the regression coefficients of any level-1 covariates into separate effects at the individual and cluster level. For example, in the application presented here, Figure 2 shows that the relationship between age 16 and age 11 scores is slightly stronger at the school-level (right panel; $r = 0.69$) than it is at the student level (left panel; $r = 0.59$). We could capture this feature of the data in our model by calculating and entering school mean age 11 score $\bar{x}_{1.j} = n_j^{-1} \sum_{i=1}^{n_j} x_{1ij}$ as an additional covariate in the model. The interpretation of the student age 11 score coefficient would then change to be a comparison of two students *in the same school* who differ in their age 11 scores by 1 SD, while the interpretation of the new school mean age 11 score coefficient would correspond to a comparison of the same student educated *in two different schools* of equal effectiveness, but which differ in their school mean age 11 scores by 1 SD. These two effects are commonly referred to as the within and contextual effects of $x_{1ij}$. The latter is often interpreted as representing a peer effect associated with being educated among higher prior attaining peers and in these data this effect would be estimated as positive.

## 3. Two-level models for longitudinal data

In this section we explore two-level models for longitudinal data. These models are essentially the same as those introduced for clustered data although the temporally ordered nature of the measurements within each individual raises new modelling challenges and possibilities. We shall focus here on the two-level random-slope model so as to build on and extend the modelling concepts presented in the previous section.

## Maths tutoring RCT data

We shall illustrate multilevel modelling of longitudinal data using simulated data for a fictitious maths tutoring randomised control trial (RCT). The data relate to an evaluation of a five-week programme aimed at accelerating student test readiness for a national maths test taken at age 11 at the end of primary schooling. The programme consists of online tutoring and structured revision that students do outside of regular schooling. The evaluation recruited a representative sample of 180 students from around England, 90 students were randomly assigned to receive the tutoring, the other 90 not. Henceforth we refer to these two groups as the treatment and control groups. All students were tested immediately before randomisation and the start of the tutoring programme and then at the end of each subsequent week. The tests were marked on a 100-point scale.

The data therefore consist of 180 students with up to six scores per student. Scores (level-1) are nested within students (level-2). Thus, in contrast to the inner-London schools example, individuals are now at level-2 in the data hierarchy. The individuals are the clusters and each individual has a collection of repeated measures associated with them. The data are unbalanced in that only 127 students (71%) have scores at all six occasions. The remaining 53 students took between 1 and 5 tests and therefore dropped out before the end of the study. Thus, we have attrition which is common feature of trials and longitudinal data more generally. Once students dropped out they were not allowed to rejoin and so the attrition is monotonic. There is no need to remove those students from the analysis who do not have the complete set of six scores. The multilevel models we fit all assume students' missing test scores are missing at random (MAR). This is a fairly reasonable assumption in the context of this application. Specifically, conditional on the test scores that we do observe for each child (and in later models treatment assignment), any subsequent missing test scores are unrelated to the unknown scores they would have obtained had they stayed in the programme. If this assumption holds then the model estimates will not be systematically biased as a result of the missingness, though they will be inefficient (less precise than if we had observed the full data).

**Model 1: Modelling individual specific trajectories in student math scores**

Before we start any modelling, it is sensible to graphically inspect the data. Figure 5 plots student math test score trajectories separately by whether they are in the treatment or control group. The plot suggests that both groups improve over time and that improvement in each case is fairly linear. There is some suggestion that students in the treatment group improve more rapidly than those in the control group. Over and above these general trends there is substantial variability in test scores both within and between students. Thus, some students start higher than others. Other students improve more rapidly than others. However, in all cases students' performances fluctuate over time with many students overtaking each other from one occasion to the next.

The simplest model which might capture most of these features is a two-level random-slope model where we regress math score on a linear time trend and where we allow every student to follow their own linear learning trajectory with a unique intercept (starting score) and slope (improvement over time). We delay allowing for a systematic difference in trajectories by treatment status until the next model.

Let $y_{ij}$ denote the math score at week $i$ ($i = 1, \ldots, n_j$) for student $j$ ($j = 1, \ldots, J$) and let $t_{ij}$ denote the time since the baseline test. We code $t_{ij}$ to range from 0 (baseline) to 1 (end of the five-week programme). Thus, with each week $t_{ij}$ increments by 0.2. This coding will prove helpful when interpreting the model. The model is written as

$$y_{ij} = \beta_0 + \beta_1 t_{ij} + u_{0j} + u_{1j} t_{ij} + e_{ij} \tag{8}$$

where $\beta_0 + \beta_1 t_{ij}$ described the overall linear learning trajectory in the data, $\beta_0 + \beta_1 t_{ij} + u_{0j} + u_{1j} t_{ij}$ is the linear learning trajectory for child $j$, and $e_{ij}$ is the residual capturing the difference between each child's observed score and that predicted by their individual trajectory. It follows that $\beta_0$ estimates the mean score at baseline across all students ($t_{ij} = 0$), while $\beta_1$ measures the mean improvement across all students by the end of programme ($t_{ij} = 1$). Thus, the coding of $t_{ij}$ leads $\beta_0$ and $\beta_1$ to estimate substantively meaningful quantities.

The random intercept effect $u_{0j}$ is the difference between child $j$'s starting performance $\beta_0 + u_{0j}$ and the overall average starting performance $\beta_0$. Similarly, the random slope effect $u_{1j}$ measures the difference between child $j$'s improvement by the end of the programme $\beta_1 + u_{1j}$ and the overall average improvement $\beta_1$.

The random intercept and slope effects are assumed bivariate normally distributed with zero mean vector and constant covariance matrix. We write these distributional assumptions as follows

$$\begin{pmatrix} u_{0j} \\ u_{1j} \end{pmatrix} \sim N \left\{ \begin{pmatrix} 0 \\ 0 \end{pmatrix}, \begin{pmatrix} \sigma_{u0}^2 & \\ \sigma_{u01} & \sigma_{u1}^2 \end{pmatrix} \right\} \tag{9}$$

where the between-student intercept variance $\sigma_{u0}^2$ measures variation in students' initial math scores and the between-student slope variance $\sigma_{u1}^2$ measures variation in students' learning gains over the duration of the programme. The between-student intercept-slope covariance $\sigma_{u01}$ measures any covariation between students' initial math scores and students' subsequent learning gains. This covariance can be reexpressed as a correlation in the usual way $\rho_{u01} = \sigma_{u01}/\sigma_{u0}\sigma_{u1}$. The level-1 residuals are assumed normally distributed with zero mean and constant variance, $e_{ij} \sim N(0, \sigma_e^2)$.

Table 2 presents the model results. The intercept $\beta_0$ and slope $\beta_1$ of the overall average line are estimated to be 54.46 and 8.39. Thus, the average student starts with a maths score of 54 percent and by the end of the course they were scoring 8.39 percentage points higher. However, here we have pooled students in the treatment and control group and so we cannot say anything about the effectiveness of the programme. That will come in the next model. The intercept and slope variances $\sigma_{u0}^2$ and $\sigma_{u1}^2$ are estimated to be 64.98 and 31.80. One way to interpret the magnitude of these parameters is to calculate the range of intercepts $(\beta_0 - 1.96\sigma_{u0}, \beta_0 + 1.96\sigma_{u0})$ and slopes $(\beta_1 - 1.96\sigma_{u1}, \beta_1 + 1.96\sigma_{u1})$ within which we expect to find the middle 95% of students in the population. Thus, the model predicts that 95% of students will start the programme with scores in the range (38.66, 70.26) and that for 95% of students their scores will improve by (-2.66, 19.44). The model therefore implies that a small minority of students will score lower at the end of the six weeks than at the beginning. The intercept-slope correlation is estimated to be 0.05 and not significant (at least according to an approximate z-test) so students' baseline scores do not appear to predict their subsequent rate of learning. Finally, the student residual variance $\sigma_e^2$ is estimated to be 22.00. Thus, at any given occasion students' scores will deviate from their learning trajectories but don't appear to do so by a particularly large amount. The 95% range in scores around their predicted values $(-1.96\sigma_e, +1.96\sigma_e)$ is estimated to be (-9.19, 9.19), so approximately plus or minus 9 percentage points.

**Model 2: Evaluating whether the intervention worked**

The above model describes the overall average student learning trajectory and student and occasion specific variation around this trend. However, our principle interest in this application is in estimating the effect of the tutoring programme on the overall average student learning trajectory. We expect there to be no difference in mean maths scores by treatment status at baseline (because students are randomly assigned to treatment), but we expect students who are tutored to learn at a faster rate over the subsequent weeks compared to students in the control group. Thus, we now extend the previous model by entering a treatment group dummy $x_j$ (0: control; 1: treatment) both as a main effect and as an interaction with time. The model can be written as follows

$$y_{ij} = \beta_0 + \beta_1 t_{ij} + \beta_2 x_j + \beta_3 t_{ij} x_j + u_{0j} + u_{1j} t_{ij} + e_{ij} \tag{10}$$

The model now separately measures the average learning trajectory in the treatment and the control group. The parameterisation of the model means that the average learning trajectory in the control group is estimated directly, while the average learning trajectory in the treatment group is estimated indirectly. To explain, in this model, $\beta_0$ and $\beta_1$ now measure the mean score at baseline and the mean increase in math scores over the six weeks *in the control group*. The coefficients $\beta_2$ and $\beta_3$ measure how much higher the mean score is at baseline and how much higher the mean increase is *in the treatment group compared to in the control group*. Thus, $\beta_0 + \beta_2$ and $\beta_1 + \beta_3$ give the mean score at baseline and the mean increase *in the treatment group*.

Table 2 presents the model results. Figure 6 plots the fitted trajectories. The treatment effect at baseline $\beta_2$ is estimated to be 0.80 and not significant ($z = 0.61$, $p = 0.544$). The treatment effect on the subsequent improvement in math scores $\beta_3$ is estimated to be 5.05 and significant ($z = 4.04$, $p < 0.005$). Thus the mean trajectory for treated students has effectively the same intercept as that for control students (54.06 vs. 54.86) but a steeper slope (10.88 vs 5.83) so that by the end of the programme students in the treatment group are scoring, on average, over five percentage points higher than their peers who did not go on the programme. The programme appears effective. Turning out attention to the random part of the model, we see that the intercept variance is effectively unchanged (64.93 vs. 64.98). This makes sense, treatment assignment cannot explain the variation in students' scores at baseline. In contrast, the slope variance reduces by 18% ($= 100(25.95 – 31.80)/31.80$). That is, 18% of the variation in score improvement seen in the data is attributable to the programme. The intercept slope correlation is again effectively zero. The residual variance is also effectively unchanged since the added covariate, treatment status, is an individual-level covariate and so cannot explain within individual variation.

Finally, it is instructive to compare the predicted trajectories from this model (Figure 6) to the observed trajectories (Figure 5). The predicted trajectories do not include the occasion specific residuals and so are straight line relationships. They can be viewed as child-specific predictions of the underlying latent learning trajectories. The plot shows both the substantial baseline variation in students' scores and the variation in students' subsequent learning rates. Indeed, many children's lines cross over one another. We also see a general fanning out of the trajectories suggesting that the variability in students' scores increases across the five occasions. Indeed, it can be shown that the model implied marginal variance (conditional on the covariates) is given by

$$\mathrm{Var}(y_{ij}|x_j, t_{ij}) = \sigma_{u0}^2 + 2\sigma_{u01}t_{ij} + \sigma_{u1}^2 t_{ij}^2 + \sigma_e^2 \tag{11}$$

Ignoring the occasion specific component $\sigma_e^2$ and substituting estimates for the remaining parameters, we use this expression to obtain the following estimates for the between-student variance at each occasion: 64.93, 66.14, 69.42, 74.78, 82.21, 91.72. We might translate these into SD or better still the range within which we expect the middle 95% of individuals to lie (see the Inner London schools example). We leave this as an exercise to the reader.

**Model 3: Allowing for autocorrelated residuals**

Recall, that in the two-level random-intercept models for the inner London schools data we calculated the ICC and VPC statistics to communicate the dependency and heterogeneity in the data. The expressions for these statistics are more complex in random slope models. They are now functions of the covariates with the random slopes, here only $t_{ij}$. Thus, the correlation between two measurements on the same subject $y_{ij}$ and $y_{i'j}$ is now a function of the timings of the two measurements $t_{ij}$ and $t_{i'j}$. This makes sense as in longitudinal data we do not expect the correlation between measurements to be constant, we expect the correlation to decay the further apart those measurements. Thus, the ICC is now given by

$$\rho \equiv \text{Corr}\big(y_{ij}, y_{i'j} \big| x_j, t_{ij}, t_{i'j}\big) = \frac{\sigma_{u0}^2 + \sigma_{u01}(t_{ij}+t_{i'j}) + \sigma_{u1}^2 t_{ij} t_{i'j}}{\sqrt{\sigma_{u0}^2 + 2\sigma_{u01} t_{ij} + \sigma_{u1}^2 t_{ij}^2 + \sigma_e^2}\sqrt{\sigma_{u0}^2 + 2\sigma_{u01} t_{i'j} + \sigma_{u1}^2 t_{i'j}^2 + \sigma_e^2}} \quad (12)$$

Substituting in the parameter estimates allows us to calculate the model-implied correlation matrix

$$\begin{pmatrix} 1.00 & & & & & \\ 0.74 & 1.00 & & & & \\ 0.73 & 0.75 & 1.00 & & & \\ 0.71 & 0.74 & 0.76 & 1.00 & & \\ 0.69 & 0.73 & 0.76 & 0.78 & 1.00 & \\ 0.66 & 0.71 & 0.74 & 0.77 & 0.79 & 1.00 \end{pmatrix} \quad (13)$$

Thus, for example, the expected correlation between the first and second occasions is 0.74, but this reduces to 0.66 when we compare the first and the last occasions. This clearly shows how the dependency in the data is no longer constant within clusters. However, the 21 correlations presented here are not freely estimated, they are derived from four parameters $\sigma_{u0}^2$, $\sigma_{u1}^2$, $\sigma_{u01}$ and $\sigma_e^2$. It is therefore sensible to compare these model-based correlations with the empirical correlations obtained by simply correlating the total residuals from the model ($r_{ij} = u_{0j} + u_{1j} t_{ij}$). These are presented below

$$\begin{pmatrix} 1.00 & & & & & \\ 0.75 & 1.00 & & & & \\ 0.74 & 0.79 & 1.00 & & & \\ 0.73 & 0.76 & 0.80 & 1.00 & & \\ 0.69 & 0.73 & 0.73 & 0.78 & 1.00 & \\ 0.63 & 0.72 & 0.64 & 0.72 & 0.78 & 1.00 \end{pmatrix} \quad (14)$$

We see that first-occasion apart correlations are in general higher, and that the five-occasion apart correlation is lower than those implied by the model. This suggests that we are not capturing the varying dependencies in the data as well as we might. It is important to correctly model the dependency in the data, not just to obtain 'correct standard errors' on the treatment effect, but also to obtain the best possible predictions for each student, a common goal in longitudinal analysis. One obvious next step is to explore the need for including a quadratic time trend by entering $t_{ij}^2$ into the model and potentially allowing this to also vary between students with the inclusion of a second random slope $u_{2j}t_{ij}^2$ and an additional slope variance $\sigma_{u2}^2$ and associated covariances $\sigma_{u02}$ and $\sigma_{u12}$.

Another possibility is to examine whether it is necessary to relax the independence assumption of the occasion specific residuals $e_{ij}$, namely $\text{Corr}(e_{ij}, e_{i'j}) = 0$. We explore this here by specifying an autoregressive order 1 structure (AR1) for the occasion-specific residuals. (Moving average and other structures are other possibilities). We can write this as

$$\text{Corr}(e_{ij}, e_{i'j}) = \rho^{|i-i'|} \tag{15}$$

where $\rho$ denotes the one-occasion apart correlation between the occasion-specific residuals. The expected correlation for two-occasions apart is then given by $\rho^2$ and so on.

Table 2 presents the results. The regression coefficients are very similar to before so we don't interpret them again. The variances, however, do differ. In particular. The intercept variance and especially the slope variance both decrease suggesting that previously we were overstating the degree of inter-individual variability in linear learning trajectories. The population is not as heterogeneous as the previous model naively suggested. We also now see a positive intercept-slope correlation, although this is not significant (at least according to a naïve z-test). In contrast, the residual variance is now larger, increasing from 21.97 to 25.18. The autocorrelation parameter is estimated to be 0.17 suggesting a certain stickiness to the occasion specific residuals. When a child performs above their long run average linear trajectory on one occasion, they are more likely than not to score above expectations at the next occasion as well. This might possibly reflect that students' true underlying performances undulate over time rather than being strictly linear. We can also think of this as reflecting unmodeled time-varying factors which persist over a limited number of occasions (e.g., stress). The autocorrelation in these occasion-specific effects, however, decays rapidly. For example, the two apart correlations is 0.03, and the , three-occasion, and further apart correlations are effectively zero. A likelihood-ratio test comparing this to the previous model confirms that this additional parameter is statistically significant ($L = 7.45$, $p = 0.006$). At this point is would be sensible to calculate again the model-implied within-student correlations and compare these to the sample correlations between the predicted total residuals, but we don't pursue this here.

**Further extensions**

The most common extensions to the two-level longitudinal model presented here is to allow for more flexible relationship between the response and time. We have already mentioned quadratic time trends. In some applications, cubic or higher order polynomials may be required. In other applications it may be more appropriate to enter time as a step function (i.e., dummy variables), or by using splines (linear or cubic), or via other more complex functions of time (e.g., with asymptotes where the outcome has a maximum such as when modelling height). In contrast, in some longitudinal settings where there is no systematic growth or development in the response over time random-intercept models with no time trend may well be sufficient.

The models presented here can be extended to include further individual characteristics into the model (e.g., gender) and to potentially interact these with time (e.g., to account for gender differences in learning rates) or other covariates (e.g., interact gender with time and treatment to explore whether the treatment effect on learning rate differs between boys and girls). Equally, one may wish to incorporate time-varying covariates to explain intra-individual variation.

We have considered here so-called 'growth curve' models. Another way to model longitudinal data is to fit so-called 'lagged-response' or 'dynamic models' where responses at previous occasions are treated as covariates. These models become relevant when there is scientific interest in establishing the direct effect of the lagged response on the current response. For example, in the current application it might be felt that doing well on the previous test will boost a student's confidence leading them to score higher on the current test.

## 4. Discussion

In this review article, we have introduced the most commonly applied multilevel models for analysing continuous responses, namely, two-level variance-component, random-intercept and random-slope models. We have illustrated the application of these models to both clustered and longitudinal data and in both observational and experimental settings. We hope that these differing examples can aid researchers in their own modelling of continuous responses. However, this is only the start of multilevel modelling. There are now a wide range of more flexible multilevel models appropriate when the data are more complex, for example in three-level and cross-classified data settings or for analysing response types other than continuous (categorical, count, and survival responses). We briefly discuss these and related extensions below, pointing to further reading in each case. We have also not yet discussed software and so also give guidance here. However, we start by reviewing classic textbooks and then online resources for those seeking more detailed and complete introductions to two-level models accompanied with wider ranges of illustrative applications than was possible to cover here.

**Textbooks**

Classic multilevel modelling textbooks aimed at the social sciences include Goldstein (2011), Hox et al. (2017), Raudenbush and Bryk (2002), Skrondal and Rabe-Hesketh (2004), and Snijders and Bosker (2012). Among these, the level of difficulty increases as we move from Hox et al. to Snijders and Bosker, to Raudenbush and Bryk, to Goldstein, to Skrondal and Rabe-Hesketh. The level of difficulty in this article is most comparable to that of Snijders and Bosker (2012). All these books cover all aspects of multilevel modelling. Good books dedicated to multilevel modelling of longitudinal data are provided by Fitzmaurice et al. (2011), Hedeker and Gibbons (2006) and Singer and Willet (2003). The difficulty in this article lies between that presented in Singer and Willett who are social science focused and that of Fitzmaurice et al. and Hedeker and Gibbons who are both medical focused. Those researching intensive longitudinal data might consult Bolger and Laurenceau (2013). For those working in education, see the edited multilevel book by O'Connell and McCoach (2008), while for those working in health see the edited multilevel model by Leyland and Goldstein (2001).

**Online course**

The Centre for Multilevel Modelling (CMM) at the University of Bristol, UK, maintain LEMMA, a free online multilevel course with over 30,000 registered users worldwide since it was launched in 2008. The course can be accessed at:

http://www.bristol.ac.uk/cmm/learning/online-course. Dedicated resources together with syntax and datasets for a range of software are available for exploring both two-level models for clustered data (Steele, 2008) and two-level models for longitudinal data (Steele, 2014) as well as many other more advanced topics in multilevel modelling.

**Software**

In terms of software, all standard packages (R, SAS, SPSS, Stata) now fit multilevel models for continuous responses. All of these software provide online documentation further describing these models. Excellent book-length treatments on multilevel modelling in these software are also available: R (Gelman and Hill, 2007), SAS (Stroup et al., 2018), SPSS (Heck et al., 2012, 2013), Stata (Rabe-Hesketh and Skrondal, 2012a, 2012b). The book by West et al. (2014) provides an overview and comparison of all these software and their options. For the analyses presented here, we have used Stata and fitted all models by maximum likelihood estimation, which is the default and standard approach in all these software. Thus, all these software will give near identical results for these models. The data and Stata software syntax to allow the reader to replicate all our results are available from the author upon request. We note that for more complex multilevel models with categorical and count responses or crossed random effects, estimation methods, computationally efficiency (speed) and therefore results can differ, sometimes appreciably, across software and so for these models we always encourage researchers to explore multiple software package, including dedicated multilevel packages such as MLwiN (Charlton et al., 2019) and HLM (Raudenbush et al., 2011).

**Three-level models**

In this article we have focused on two-level settings where, for example, individuals at level-

1 are nested within clusters at level-2 and where we include cluster random effects to account for the variability in the response and the regression coefficients across clusters. In many applications the clusters may themselves be further nested within superclusters. Rasbash et al. (2010) present an application similar to our inner-London schools example, but where the schools are further nested within school districts. The authors were interested in the importance of school district effects relative to school effects. The former only accounted for around 2% of the variation in student progress whereas schools accounted for around 12%. So once again, students accounted for the biggest share in variation, some 86%. Once one has the hang of two-level models, extending these models to incorporate a third or even a fourth or higher levels of clustering is relatively straightforward. Indeed, an interesting extension to their analysis would have been to insert teachers as a further level situated between the student and school level. We would simply introduce an additional random intercept defined at this new level. Where appropriate, we can then additionally include random slopes at any given higher level on covariates defined at any lower level. See Leckie, 2013c for a dedicated introduction to the concepts and practice of three-level models including software syntax and data to replicate all examples.

**Cross-classified models**

In many studies, however, the data will not follow a strict hierarchy. Rather the observations or level-1 units are best described as being separately and simultaneously nested within two different classifications of clusters, both conceptually at level-2. Leckie and Baird (2010) present an application to raters scoring exemplar student essays as part of a national rater benchmarking (calibration) exercise prior to live scoring of real student essays in standardised tests. Each exemplar essay was scored by multiple raters and each rater scored multiple essays. The data are therefore cross-classified with the assigned scores (level-1) nested separately and simultaneously within students (level-2) and raters (also conceptually at level-2). The assigned scores were then compared to the score assigned by the chief examiner (the gold standard to which raters were attempting to align). Leckie and Baird fitted cross-classified models to these score differences to estimate rater effects on student scores (severity/leniency) having adjusted for the different mix of essays scored by each rater (student ability). Thus, the model included separate rater and student random effects, but where neither is declared to be nested within the other. They found raters varied in the severity with which they scored with greater variation exhibited among less experienced raters. See Leckie, 2013a for a dedicated introduction to the concepts and practice of cross-classified models including software syntax and data to replicate all examples. A formal study of the impact of ignoring cross-classified structures is provided by Meyers and Beretvas (2006).

**Multiple membership models**

We have described cross-classified data as one example of how standard data hierarchies can break down. Another example is when multiple membership structures are present and rather than each individual belonging to a single cluster some or many individuals may simultaneously belong to multiple clusters. Leckie (2009) present an application similar to our inner-London schools example, but where we observe not just the final school in which students took their age 16 exams, but where children change schools over the course of secondary schooling. One should clearly account for the full sequence of schools attended, especially when students move just before their examinations. Leckie does this by replacing the usual school random effect in a two-level random-intercept model with a weighted sum of all the school random effects where these weights represent the relative membership (contribution or influence) of each school on each student. For example, where a student attends two schools, the weights for these two schools will reflect the proportion of time spent in each school while the weights for all other schools will be set equal to zero. See Leckie, 2013b for a dedicated introduction to the concepts and practice of multiple membership models including software syntax and data to replicate all examples. A formal study of the impact of ignoring multiple membership structures is provided by Chung and Beretvas (2012).

**Multivariate response models**

We have focused exclusively on multilevel models for analysing single responses. However, in some applications interest lies in simultaneously studying multiple outcome variables and how the outcome-covariate relationships vary across these responses. When the data are clustered, we can extend conventional multiple response equation models to include a separate cluster random effect in each equation. These random effects are then correlated across equations as typically are the residual errors. Thus, interest lies in studying the strength of these cross-equation correlations at each level of analysis and how they change as further covariates are added to the model. For example, Leckie (2018) fitted a two-level students-within-school random-intercept bivariate continuous response model to student achievement in English and maths. The model adjusted for student prior achievement and so the random effects were interpreted as value-added school effects. The school-level correlation between these random effects was estimated to be 0.786 indicating that schools that were effective in one subjected tended, but were not guaranteed, to be effective in the other subject. Schools were considered to be relatively consistent in their influence on student achievement across these two academic subjects.

**Categorical responses, counts, and survival**

Finally**,** we have focused exclusively on multilevel modelling of continuous responses. However, we can equally fit multilevel models to categorical responses (binary, ordinal, nominal), counts and survival data. Thus, we can extend, for example, conventional logistic and Poisson regression for binary and count responses to both include cluster random effects. Many of the concepts introduced here (variance component model, random intercept model, random coefficient model, ICC, VPC, predicted random effects) carry over, but the analysis of these more complex response types introduce a range of new interpretation, estimation and software challenges which are important for users to be familiar with.

Table 1.

Estimates for inner-London schools exam scores data

|  |  | Model 1 | | Model 2 | | Model 3 | |
|---|---|---|---|---|---|---|---|
|  |  | Est | SE | Est | SE | Est | SE |
| $\beta_0$ | Intercept | -0.013 | 0.054 | 0.002 | 0.040 | -0.168 | 0.054 |
| $\beta_1$ | Age 11 scores | | | 0.563 | 0.012 | 0.560 | 0.012 |
| $\beta_2$ | Girl | | | | | 0.167 | 0.034 |
| $\beta_3$ | Boys' school | | | | | 0.178 | 0.111 |
| $\beta_4$ | Girls' school | | | | | 0.159 | 0.087 |
| $\sigma_u^2$ | School variance | 0.169 | 0.033 | 0.092 | 0.019 | 0.081 | 0.017 |
| $\sigma_e^2$ | Student variance | 0.848 | 0.019 | 0.566 | 0.013 | 0.562 | 0.013 |
| $\sigma_r^2$ | Total variance | 1.017 | | 0.658 | | 0.643 | |
| $\rho$ | ICC/VPC | 0.166 | | 0.140 | | 0.126 | |
| $D$ | Deviance | 11010 | | 9357 | | 9325 | |

Note.

The response is student age 16 score. $N = 4059$ students nested within $J = 65$ schools.

Table 2.

Estimates for maths tutoring RCT data

| | | Model 1 | | Model 2 | | Model 3 | |
|---|---|---|---|---|---|---|---|
| | | Est | SE | Est | SE | Est | SE |
| $\beta_0$ | Intercept | 54.46 | 0.66 | 54.06 | 0.93 | 54.09 | 0.93 |
| $\beta_1$ | Time | 8.39 | 0.65 | 5.83 | 0.89 | 5.83 | 0.89 |
| $\beta_2$ | Treatment | | | 0.80 | 1.31 | 0.76 | 1.31 |
| $\beta_3$ | Treatment $\times$ Time | | | 5.05 | 1.25 | 5.06 | 1.25 |
| $\sigma_{u0}^2$ | Intercept variance | 64.98 | 8.18 | 64.93 | 8.17 | 60.85 | 8.48 |
| $\sigma_{u1}^2$ | Slope variance | 31.80 | 7.78 | 25.95 | 7.09 | 15.60 | 8.69 |
| $\sigma_{u01}$ | Intercept-slope covariance | 2.07 | 5.87 | 0.42 | 5.64 | 4.77 | 6.10 |
| $\rho_{u01}$ | Intercept-slope correlation | 0.05 | 0.13 | 0.01 | 0.14 | 0.15 | 0.23 |
| $\sigma_e^2$ | Residual variance | 22.00 | 1.26 | 21.97 | 1.25 | 25.18 | 2.26 |
| $\rho$ | Residual autocorrelation | | | | | 0.17 | 0.07 |
| $D$ | Deviance | 6229 | | 6211 | | 6203 | |

Note.

The response is student math score. $N = 948$ measurements nested within $J = 180$ children.

Figure 1.

Student and school mean age 16 scores plotted by school.

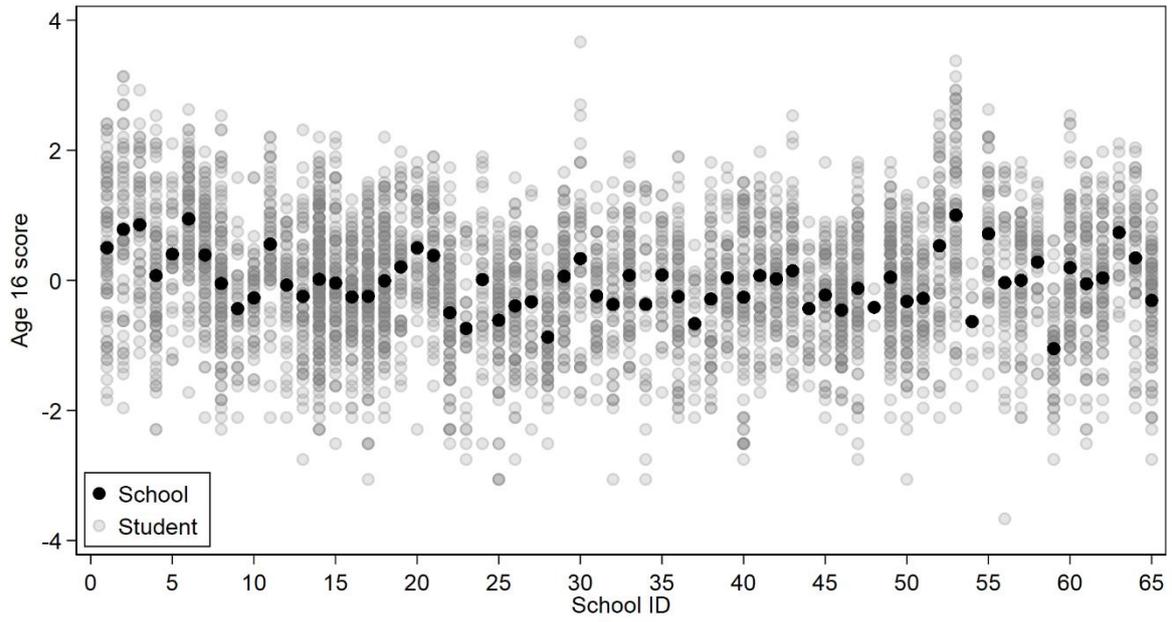

Figure 2.

Scatterplots of student age 16 scores against age 11 scores (left plot) and school average age

16 scores against age 11 scores (right plot). Pearson correlations are reported.

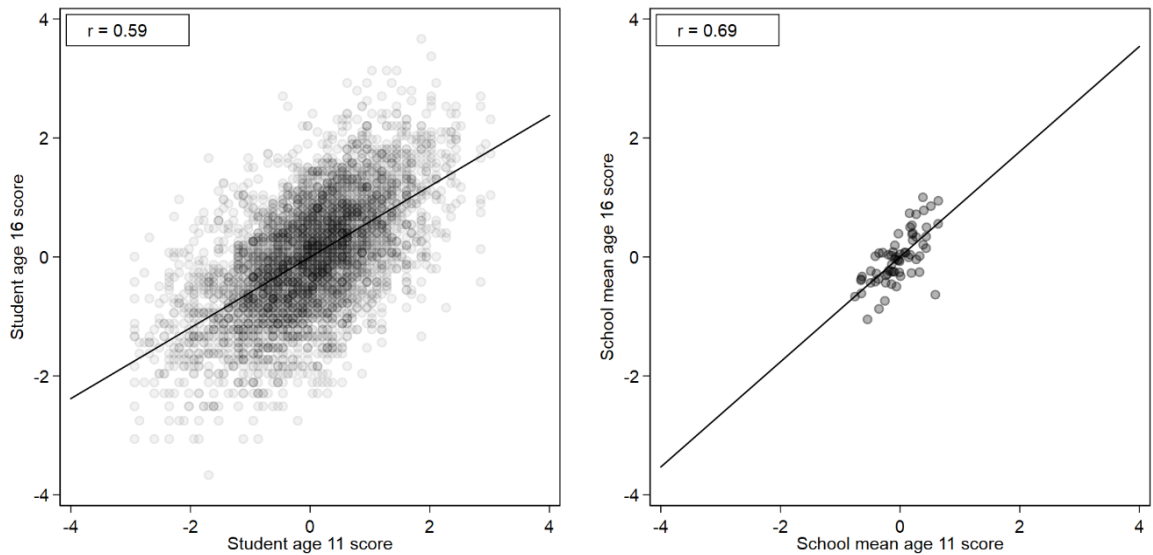

Figure 3.

Predicted school lines based on model 2 (left) and a random-slope version of that model

(right)

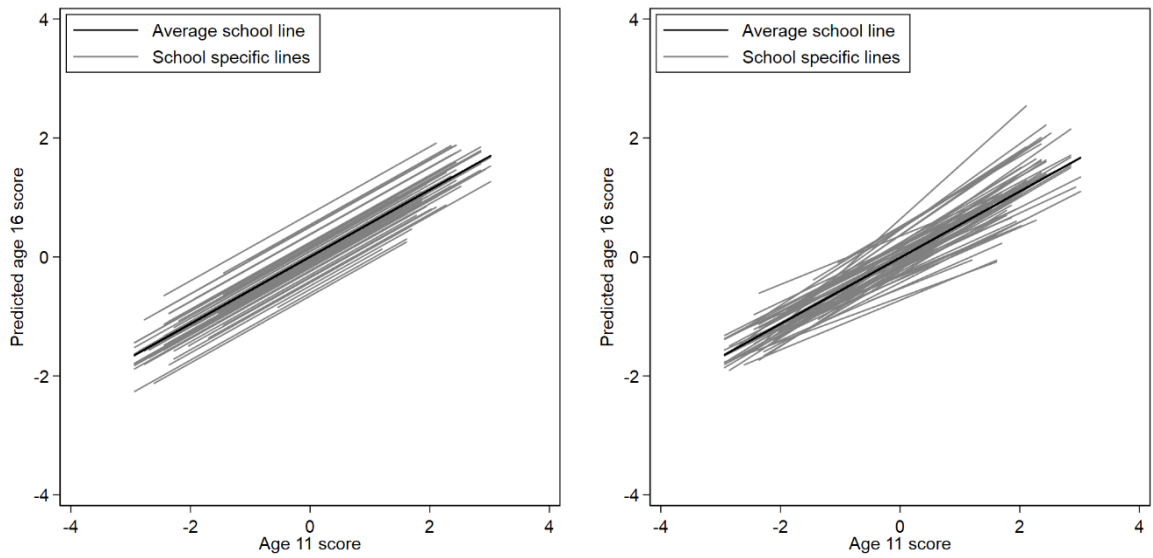

Figure 4.

Caterpillar plot of the predicted school effects based on model 2. Point estimates are plotted with 95% confidence intervals. School IDs are superimposed.

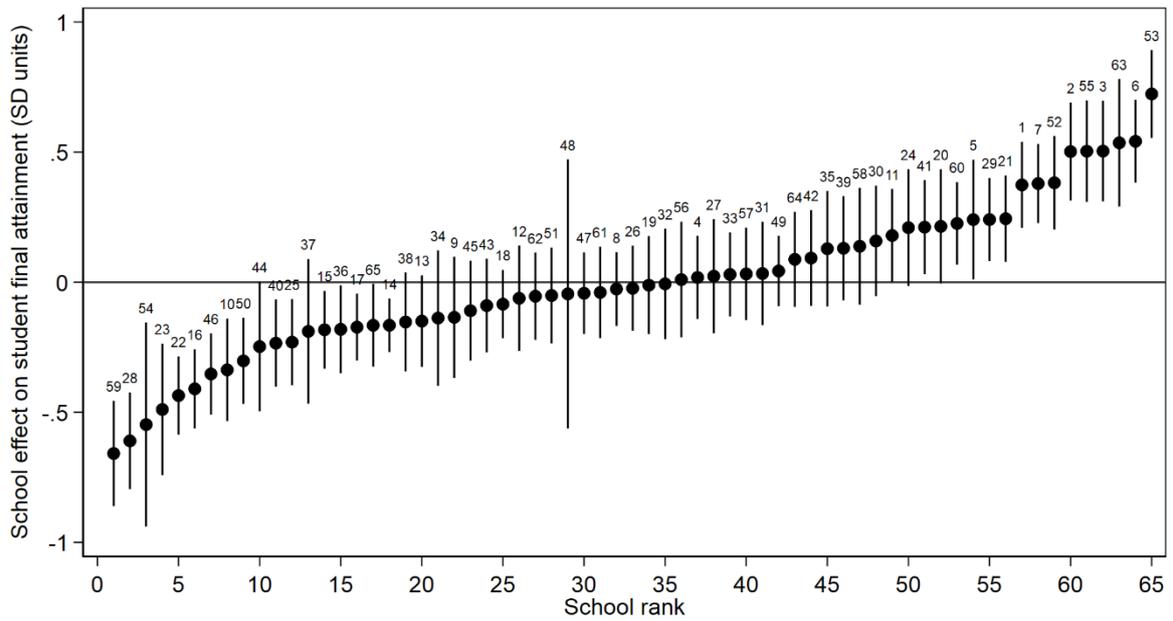

Figure 5.

Observed average and student-specific learning trajectories by treatment group

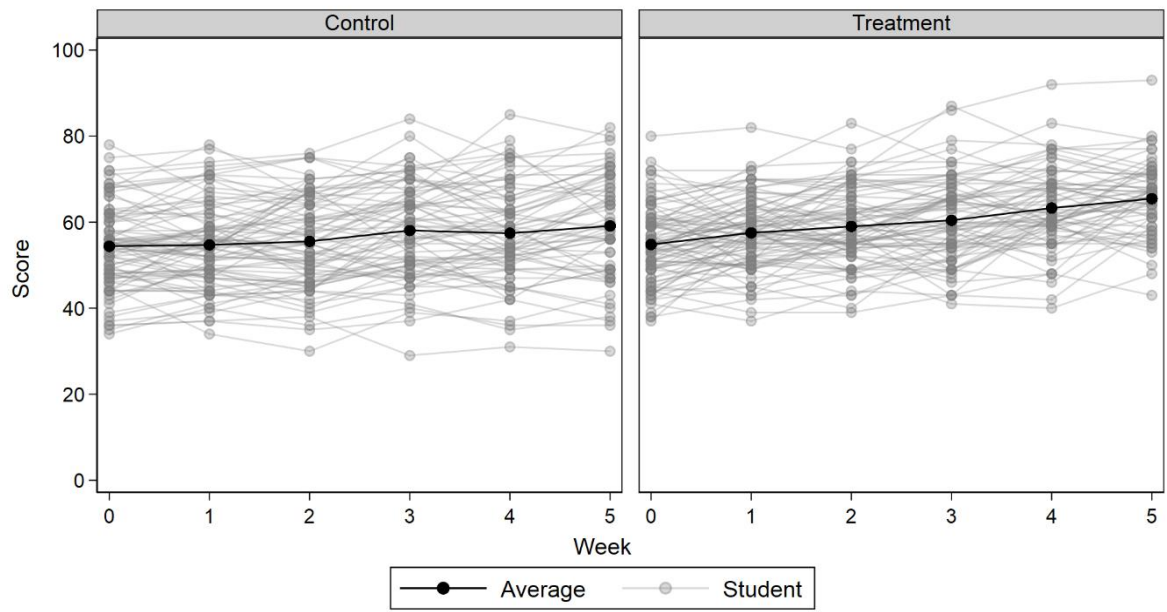

Figure 6.

Predicted population-averaged and student-specific learning trajectories by treatment group.

Predictions based on model 2.

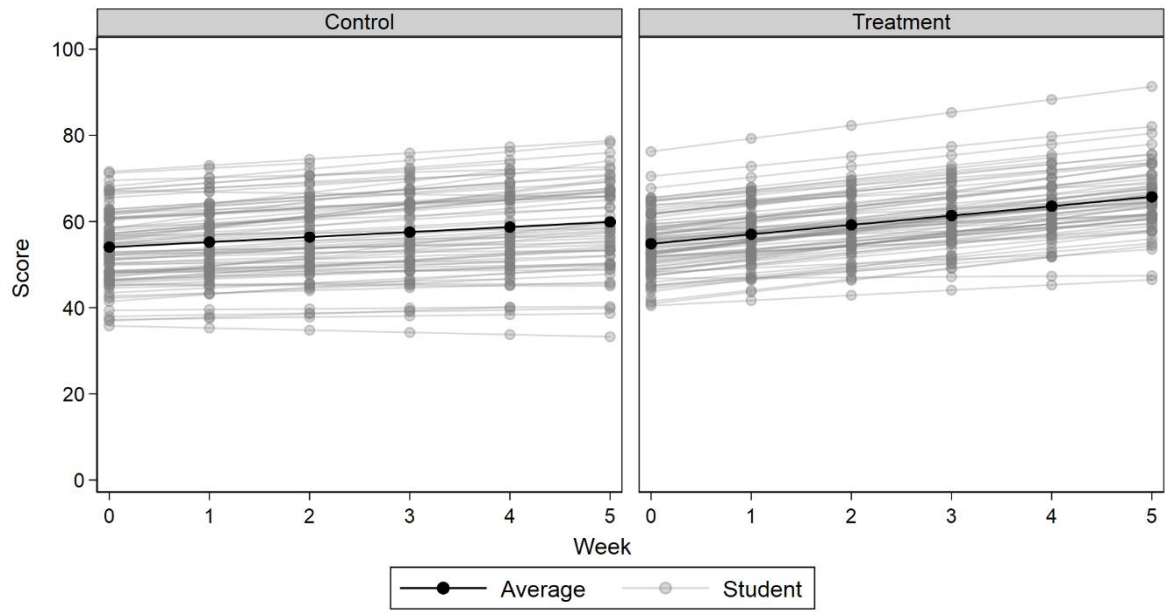